\documentstyle[12pt,aasms4]{article} \newcommand{\etal}{{\it et al. \,}}

\begin{document}

\title{The Star Formation History of the Local Group Dwarf Elliptical
Galaxy NGC 185. II. Gradients in the Stellar Population
\altaffilmark{1}}

\author{D. Mart\'\i nez-Delgado \altaffilmark{2} and A. Aparicio}

\affil{Instituto de Astrof\'\i sica de Canarias, E-38200 La Laguna,
Canary Islands, Spain\\ Electronic mail: ddelgado@iac.es, aaj@iac.es}

\and

\author{C. Gallart\altaffilmark{3}}

\affil{Department of Astronomy, Yale University, P.O. Box 208101,
New Haven, Connecticut 06520, USA and Departamento de Astronom\'\i
a, Universidad de Chile, Casilla 36-D, Santiago, Chile \\Electronic
mail: carme@astro.yale.edu}

\altaffiltext{1} {Based on observations made with the William Herschel
Telescope operated on the island of La Palma by the Isaac Newton Group
and with the 2.5 m Nordic Optical Telescope operated by NOT S.A. at
the Spanish Observatorio del Roque de los Muchachos of the Instituto de
Astrof\'\i sica de Canarias.} \altaffiltext{2} {Guest observer of the
ING Data Centre, Cambridge, UK.} \altaffiltext{3} {Yale Andes Fellow.}

\begin{abstract}

 The star formation history of the dE NGC 185, together with
its spatial variations,  has been investigated for old, intermediate-age,
 and
young stars using new ground-based $H_\alpha$ and $BVI$ photometry, and
synthetic color--magnitude diagrams (CMDs). We find that the bulk of the
stars were formed in NGC 185 at an early epoch of its evolution. After
that, the star formation proceeded at a low rate until the recent past,
the age of the most recent traces of star formation activity detected
in the galaxy being some 100 Myr.

As for the spatial variations, the star formation rate, $\psi(t)$,
for old and intermediate ages shows a gradient in the sense of taking
smaller values for higher galactocentric radii. Moreover, recent star
formation is detected in the central $150 \times 90$ pc$^2$ only, where the
youngest, 100 Myr old population is found. No traces of stars born more
recently than 1 Gyr ago are found outside this central region. Since
the larger concentration of stars of any age lies in the central part
of a galaxy, it could be the case that the youngest stars originate
from material ejected from  dying stars and that this process would
only be efficient enough in the center of the galaxy.

The luminous blue {\it stars} discovered by Baade (1951) in the center of
NGC 185 are discussed using new  CCD images in $B$ and Baade's  original
photographic plates. Considering their fuzzy, unresolved appearance and
that a conspicuous main sequence is lacking in the CMD at our limiting
magnitude, we reach the conclusion that most of Baade's blue objects
are in fact star clusters. These clusters, as well as the other stellar
populations, are young (a few 100 Myr), but not as much as they would be
if they were individual stars (a few 10 Myr).

A supernova (SN) remnant close to the center of NGC 185 has been analyzed
from $H_\alpha$ images. The fact that a conspicuous MS is lacking in our
CMD implies that the SN had  originated from a white dwarf progenitor.

A consistent picture arises in which the gas observed in the central
region of NGC 185 would have an internal origin. The rate at which
evolved stars return gas to the ISM is enough to seed the recent star
formation observed in the center of the galaxy and the SN rate is
probably low enough to allow the galaxy to retain the gas not used in
the new stellar generations. Further support is found in the similar
kinematical properties of gas and stars.

\end{abstract}


\section{Introduction} \label{intro}

From the early work of Baade (1944a, 1944b), who resolved the four
Andromeda dE companions into individual stars, elliptical galaxies were
considered to be essentially old, coeval systems with ages comparable
to those of  Milky Way population II globular clusters. In the light of recent
data, however, there is evidence that the majority of the Local Group
dE galaxies have undergone recent star formation activity.  The
study of the stellar content of dE galaxies by means of their 
color--magnitude diagrams (CMDs) provides the most direct method of establishing
whether they have had star formation episodes since the initial primeval
event and even to locate in time, in a more precise way, that initial
star formation event.  This is the first step in correctly interpreting
their integrated light by means of spectral synthesis techniques, which
is the only information available at large distances.  The Local Group dE
galaxies offer a unique opportunity to study their evolution in detail
by this means.

NGC 185 is a dE companion of the Andromeda galaxy. The presence of a dozen
of bright, blue stars and two conspicuous dust patches in the central
area of NGC 185 was firstly noted by Baade (1951). These ``population I''
features indicated that NGC 185 did not fit the concept of dE galaxies
as pure population II systems and, for this reason, it was classified
by Sandage \& Tammann (1981) as a peculiar dE (dE3p). The possibility
that a young population might exist in the center of NGC 185 was also
discussed by Hodge (1963), who showed that the galaxy is bluer in the
center. More recently Saha \& Hoessel (1990) surveyed the variable stars,
detecting about 150 RR Lyrae stars, which implies the presence of an old stellar
population (older than some 10 Gyr). Recent CMDs (Lee, Freedman, \& Madore
1993; Mart\'\i nez-Delgado \& Aparicio 1998, Paper I) showed that its
bright stellar population is dominated by red giant branch (RGB) stars
and a significant number of luminous red stars above the tip of the RGB
(TRGB). The latter could be an intermediate-age asymptotic giant branch
(AGB) population and, together with the aforementioned population I
traces, would suggest that  star formation in NGC 185 has occurred
over an extend period of time.

We thus have a picture of an object presumably dominated by an old
population, but showing evidence of recent star formation activity. We
present here the analysis of its resolved stellar content using the method
based on synthetic CMDs that we have employed to study the SFHs of other
Local Group dwarf galaxies (Gallart {\it et al.} 1996a,b; Aparicio,
Gallart \& Bertelli 1997a,b; Gallart {\it et al.} 1999). It provides a
better insight into the relative contribution of the stellar population
at different ages and a more complete view of its SFH.

In this paper, the SFH of NGC 185 is discussed by means of model CMDs
in the light of new ground-based $BVI$ CCD photometry. The wide field
covered by our images allows also to study the gradients in the SFH as a
function of galactocentric radius, which can yield important information
to constrain dwarf galaxy formation models. In Sec. \ref{observation}
we present the observations and data reduction; in Sec. \ref{young}
we discuss the young stellar population. In Sec. \ref{sfh} we
present our derivation of the SFH, based on synthetic CMDs, and takes
into account the information provided by the bright red stars above
the TRGB. In Sec. \ref{gas} we discuss the origin of the young population and the gas in NGC 185. Sec. \ref{global} is devoted to the global properties of
NGC 185. Finally, in Sec.  \ref{conclusion} we summarize the results
presented throughout the paper.

\section{Observations and data reduction} \label{observation}

The observational $VI$ data of NGC 185 have been discussed in detail in
Paper I. In addition, images of its central region were obtained with a
Johnson $B$ filter and with a narrow band $H_\alpha$ filter using the
2.5 m Nordic Optical Telescope (NOT)
 at the Roque de Los Muchachos Observatory on the island of La Palma. The
HiRAC camera with a $2048\times 2048$ Loral CCD binned to $2\times2$
was used. After binning, it provides a scale of 0.22 $''$/pix and a total
field of $3.75\times 3.75$ ($'$)$^2$. Total integration times were 900 s
in $B$ and 1800 s in $H_\alpha$. A 900 s exposure was also taken with
an $H_\alpha$-continuum filter. Moreover, 900 s $B$, 600 s $V$ and 400
s $I$ images were obtained of a nearby field situated 25$\arcmin$ north
of the center of NGC 185 to correct the foreground contamination of the
broad-band images. Table 1 lists the NOT journal of observations. It has
to be added to the journal of observations given in Paper I. The seeing
during this campaign was excellent, with values of $\sim$ 0.5 $\arcsec$
during good part of the nights.

Bias and flat-field corrections were carried out with IRAF. DAOPHOT and ALLSTAR
(Stetson 1994) were then used to obtain the instrumental photometry of
the stars. Eighteen standard stars from the list of Landolt (1992) were
measured during the observing run to calculate the atmospheric extinctions
for each night and the equations transforming into the Johnson-Cousins
standard photometric system. A total of about 180 measures in each band
a($B$, $V$, and $I$) were used. The transformation equations are:

\begin{equation} (B-b)=25.647+0.023(B-V); ~~~~\sigma=0.006 \end{equation}
\begin{equation} (V-v)=25.251-0.103(V-I); ~~~~\sigma=0.005 \end{equation}
\begin{equation} (I-i)=24.557+0.007(V-I); ~~~~\sigma=0.006 \end{equation}

\noindent where capital letters stand for Johnson--Cousins magnitudes
and lower-case letters refer to instrumental magnitudes corrected for
atmospheric extinction. The $\sigma$ values are the dispersions of
the fits at the barycenters of the point distribution; hence the
internal zero-point errors are minimal. The dispersions of the
extinction for each night vary from $\sigma=0.009$ to $\sigma=0.019$,
 and the error of the aperture corrections are of the order of 0.03. Putting all these
values together, the total zero-point error of our photometry is
estimated to be about 0.04  for each band.

 This calibration was used to transform our instrumental magnitudes of
the comparison field into the Johnson--Cousins standard system. However,
for consistency with Paper I, the $B$ photometry of the central field
of NGC 185 was calibrated using photometric data by Lee {\it et al.}
(1993). Figure \ref{calibraB} shows the residuals between the magnitude
of Lee {\it et al.} minus our instrumental magnitude, ($B_{\rm Lee} - b$),
versus Lee {\it et al.} magnitude.  The slope of the transformation and
the color term is comparable to its dispersion, so we adopted the mean
value, $B_{\rm Lee} - b=25.163 \pm 0.004$ to transform to Johnson--Cousins
magnitudes. In addition, one standard star from Oke (1990) was observed
three times to calibrate the $H\alpha$ images. The dispersion of the
measures was less than 0.01 mag.

Finally, artificial star tests for the comparison field frames and for the
$B$ image of NGC 185 were performed in a  similar way to that described
in Paper I to obtain the {\it crowding-trial} table (see Aparicio \&
Gallart 1995), from which observational effects can be simulated in the
model CMDs.

\section{Recent star formation in the central region of NGC
185}\label{young}

\subsection{The resolved young stellar population}\label{resolved}

Since reported by Baade (1951), the existence of several blue
stars in the center of NGC 185, indicating a young stellar population,
has been an intriguing feature of this galaxy. Hodge (1963) showed
that NGC 185 is bluer in the center and that the twelve OB-type stars
discovered by Baade (1951) cannot alone explain the observed blue color,
suggesting that they merely are the brightest members of a population I
component of $\sim 2 \times 10^{5}$ M$_{\sun}$. The photometry of this resolved blue population were published for first time by Lee {\it et al.} (1993). Blue stars were also reported by Lee {\it et al.} (1993) who for the first time published photometry of these objects. Kim \& Lee (1998) found
that the color of NGC 185 gets inwardly bluer  at $R < 25 \arcsec$ while
it remains constant outside this radius. These indications of recent
star formation in the center of NGC 185 are also in agreement with the
image shown in Figure \ref{color}. This image is  composed of $B$,
$V$, and $R$ images obtained with the 1 m Jacobus Kapteyn Telescope (JKT)
 at the Roque de los Muchachos
Observatory, and retrieved from the archive of the Isaac Newton
Group. The blue region is confined in the central $\sim 50\arcsec
\times 30\arcsec$ of the galaxy, which corresponds to 150 $\times$
90 pc$^2$ at the distance of NGC 185 (616 kpc).

Unfortunately, the study of the resolved young component of NGC 185
from ground-based telescopes is extremely limited by crowding, even under
good seeing conditions. Fig. \ref{cmd} shows the $[(B-V),V]$ CMD of
the central field of NGC 185 ($3.75\arcmin \times 3.75 \arcmin$). The
main structure is the {\it red tangle} (see Aparicio \& Gallart 1994)
populated by RGB and AGB stars. Besides this, a small number of blue
and yellow stars are also observed with $(B-V)<1.2$, that might be the
trace of recent star formation activity. Figure \ref{cmdcomp} shows the
CMD of a companion comparison field populated by stars in the
range $0.4<(B-V)<1.4$, indicating that many of the possible young stars
in the CMD of NGC 185 are probably foreground stars.

To remove the effects of the foreground contamination, we have derived
the $V$ luminosity function (LF) of the galaxy and the comparison field
for stars with $0.0<(B-V)<1.5$ and $17$ mag $<V<$ 23 mag. To select blue stars, only the stars bluer than the line defined by $V=-2.2 (B-V) +24.3$ and $V< 21$ has been considered. The star counts have been corrected of completeness using completeness
factors $\Lambda$ for the central region of the galaxy derived from
the {\it crowding-trial} table discussed in Section \ref{observation}.
Figure \ref{lf} shows the resulting completeness-corrected LFs for NGC 185
(solid line) and the foreground field (dotted line). While the stars with $V<20.5$ mag in the NGC 185 field can be
accounted for by foreground contamination, a significant excess of
stars is observed in the range $20.5$ mag $<V<$ 23 mag.  It is mainly 
produced by
the clump of stars at $(B-V)\sim 1.0$ and $V\sim 21.5$ mag  (Fig. \ref{cmd}). These stars may be young, core He-burning
(HeB) intermediate-mass stars. The lack of information about the metal
abundance of the young population of NGC 185 precludes an accurate
determination of the age of these stars, but an approximate estimate
can still be given. In Paper I, the maximum metallicity of the RGB
stars was estimated to be [Fe/H] $\sim -1.1$ or $Z\sim 0.001$.
Assuming a monotonic chemical enrichment, this should be a lower
limit to the metallicity of the young population in the center of the
galaxy. Therefore, assuming $0.001<Z<0.01$ for the young stars and using
theoretical isochrones from the Padua library (Bertelli \etal 1994,
and references therein), we obtain  an age in the range 100--125 Myr
for these stars. The $Z=0.001$ isochrone and age 125 Myr is plotted
in Figure \ref{cmd_blue}.

 Is the upper main sequence (MS) of a population of such an age
visible in the lower, blue part of the CMD? To check this, a synthetic
CMD has been computed with an  SFR running from 15 
to 125 Myr ago. The observational errors corresponding to the central
region of NGC 185 have been simulated to produce what we will
term a {\it model} CMD. Both, synthetic and model CMDs are plotted in
Figure \ref{cm125}. It shows that the age of 125 Myr is consistent with
conspicuous traces of the MS being absent in the CMD. If the former age
estimate is correct, deeper, well sampled CMDs of the center of NGC
185 should show traces of an MS reaching $V\sim 23$ to 23.5 mag.

The CMD of the comparison field  also sheds light on the nature of the
yellow stars reported in Paper I populating a strip around $[(V-I),I]\sim
[0.6,21]$ to [1.1,17]. The possibility was discussed there that
they could be red supergiants coming from recent star formation extended
over the whole galaxy. The total number of these stars found in the NGC
185 field was 53, quite more than the 27 foreground stars predicted by the
Bahcall \& Soneira (1980) model of the Galaxy. However, using the counts
from our comparison field after scaling to the area of the NGC 185 frame,
74 foreground stars are predicted in that region of the CMD, ruling out
the possibility that significant star formation had taken place in the
last 1 Gyr outside the inner region of NGC 185.

\subsection{Baade's blue stars} \label{baade}

The dozen of luminous, blue stars in the center of NGC 185, with
photographic magnitudes between 20 and 21, reported by Baade (1951), have been
interpreted as impurities in a predominantly type II population. The
discovery was made on photographic plates taken with the 5 m Hale
Telescope at Mount Palomar (California, USA), but a finding chart was
never published. These stars have played a relevant role in the
evolution from the {\it pure population-II} concept of dE galaxies to the
present view of these systems as populated in general by a mixture of
old, intermediate-age and young stars. For this reason,  we thought it
would be interesting to go back to Baade's original photographic plates, and
Fig. \ref{placa} reproduces one of them. It must be noted that what Baade intented to say by "blue" was that the stars were brighter in the blue than in the red plates and, in any term, bluer than the general population in the field. We maintain the term "blue" here for consistency with Baade work.

From visual inspection and a
comparison of the blue and red plates, we have selected 19 blue stars in the
central region of NGC 185, among which the dozen blue stars mentioned by
Baade should hopefully be included. It is worth noting that, also from
visual inspection of the original Baade's plates, Hodge (1973) found
17 population I stars close to the center of the galaxy. The finding
chart for the 19 stars we have selected is shown in Figure \ref{map}. The
positions (in pixels) of the stars in the frame and their $BVI$ photometry 
are listed
in Table 2. Their cross-identifications with those listed
by Lee {\it et al.} (1993) (their Table 3) are given in column 2.  The stars are predominantly located in the central $\sim
20''$ ($\sim$ 60 pc) of the galaxy, within the bluer central region shown
in Figure \ref{color}. Only one foreground star is expected in this area
and in the corresponding region of the CMD, which confirms the idea that
they  belong to NGC 185.

The  actual nature of these objects, marked
with filled dots and pentagons in Fig. \ref{cmd_blue}, is puzzling.
 Their location in
the CMD is compatible with that of intermediate to massive, evolved He-burning
stars with ages between 40 and 150 Myr (for the assumed $0.001<Z<0.01$
metallicity range), which is in agreement with the estimate by Lee {\it
et al.} (1993) of the minimum age for the blue/yellow population of $\sim$
 20--40 Myr from the brighter stars lying along the ZAMS line. However,
such a population requires the presence of an MS counterpart, which is
not visible. This fact is shown in Fig. \ref{cm60} where a synthetic CMD and
its corresponding crowding-simulated model CMD for a stellar system with
constant SFR from 1 to 40 Myr ago are plotted. Together
with the MS, the model CMD of Fig. \ref{cm60} shows that some bright
core He-burning stars should also be observed at $V\sim$ 21 mag and $(B-V)\sim$ 1.4. Due
to the metallicity dependency of the color indices of the bright, blue,
HeB stars, less relevance has to be given to the fact that they show
a diagonal sequence in the model CMD rather than a fuzzy distribution
like that in the observational CMD.

 Another possible interpretation of Baade's stars is that they could
be in fact be multiple stars. Our best images, with seeing $\sim 0.6
\arcsec$, reveal that some of these blue objects show a fuzzy aspect and
larger FWHMs. The most clear examples are objects  3 and 12 in Table
2 (see also Fig. \ref{map}), which show clear non-stellar appearances with
a FWHM $> 1.5\arcsec$. At the distance of NGC 185, this corresponds to
a linear size of $\sim 5$ pc, similar to typical sizes of Galactic open
clusters, like the Pleiades or Praesepe (4 pc). Also their absolute magnitudes
and colors are in the range of values of Galactic open cluster (Sagar,
Joshi, \& Sinvhal 1983). For comparison, the colors and magnitudes of
the Pleiades, Praesepe and NGC 188 as representative of young, intermediate-age,
and old open clusters, have been plotted in Figure \ref{cmd_blue}.

These star clusters could be similar to those found in NGC 205 (Hodge
3, Hodge 1973; Hubble V, Da Costa \& Mould 1988; Geisler {\it et al.} 1999), although fainter. Furthermore, Peletier (1993) and Lee (1996) has identified some bright blue objects in the central region of NGC 205, which recent {\it HST} images have shown to be actually stellar clusters (Cappellari {\it et al.} 1999; Geisler, private communication).

All these pieces of information lead us to conclude that most or
all the bright blue objects found by Baade in NGC 185, are young star
clusters and associations. They must be young, to account for their
blue colors, but the fact that they are probably not single stars results
in greater ages and lower star formation rates (SFRs). In particular,
the minimum age can be estimated to be about 100 Myr, consistent with
our estimated age of the latest star formation activity in NGC 185
(Section 3.1).

\subsection{The star formation rate for the last 1 Gyr} \label{sfryoung}

The results on the number of blue and yellow stars in the CMD can be
used to estimate the average SFR for the last 1 Gyr in NGC 185. This
is possible due to the distributions of stars older and younger than
$\sim 1$ Gyr being decoupled in the CMD: almost no stars older than 1
Gyr are among the blue population and almost no stars younger than that
age populate the red tangle (see Gallart {\it et al.} 1996a).

With this aim, after subtracting the foreground star contribution,
we integrated the crowding-corrected LF obtained in Sec. \ref{resolved}
to estimate the total number of blue/yellow stars. This number was then
scaled to that observed in a model CMD computed with constant star
formation from 1 Gyr to 125 Myr ago and with metallicities randomly
distributed in the range $0.001<Z<0.01$, following a method similar to
that explained in Aparicio {\it et al.} (1997c). The resulting SFR is
$\bar{\psi}_{1-0}=6.6\times 10^{-4}$ M$_{\sun}$ yr$^{-1}$. This would be
the SFR if all the blue/yellow objects were actually stars and must hence
be considered as an upper value and, in any case, as a rough estimate,
since no confident estimate can be made of how many of the  brightest blue
objects are single stars.

\subsection{The emission nebula in the center of NGC 185}\label{snr}

Gallagher, Hunter, \& Mould (1984) discovered an emission nebula near
the core of NGC 185. They measured  [SII]:$H_{\alpha}=1.2\pm 0.3$ from which they concluded that
it was probably  a supernova remnant (SNR). Ho, Filippenko, \& Sargent
(1997) confirmed this result ([SII]$:H_{\alpha}=1.5$), but classified
the NGC 185's nucleus as Seyfert 2. Young \& Lo (1997) obtained the
first $H_{\alpha}$ image of this nebula and found that its properties are
compatible with those of SNRs observed in the Local Group. Finally, it
has  been detected neither in the VLA radio search by Dickel, D'odorico,
\& Silverman (1985) nor in the recent {\it ROSAT} HRI X-ray observations by
Brandt {\it et al.}  (1997), suggesting that it could be old (Mateo 1998).

We have obtained new $H_{\alpha}$ observations of the central region
of NGC 185 under good seeing conditions ($\sim 0.7\arcsec$) that
show the emission nebula in more detail. Figure \ref{halfa} shows the
continuum-subtracted $H_{\alpha}$ image. It is quite extended and has an
arc-like morphology, suggesting that it may be a portion of a larger,
old remnant. Its center is close to the nucleus of the galaxy. In
addition to the nebula, two compact $H_{\alpha}$ objects can be seen
in Fig. \ref{halfa} close to the center of the galaxy. They are likely
planetary nebulae and, to our knowledge,  had not been previously
identified. Fitting an ellipse to the shape of the partial shell of the
diffuse nebula, we have estimated a diameter ($D$) of 80 pc as the mean of
the major and minor axes. We have measured a flux of $H_{\alpha}=3.0\times
10^{-14}$ erg s$^{-1}$ cm$^{-2}$ in very good agreement with Young \&
Lo (1997). For a distance of 616 kpc (Paper I), the luminosity results
$L_{H\alpha}=1.3\times 10^{36}$erg s$^{-1}$.

The evolution of an SNR is dominated by its interaction with the
ISM. Although this precludes accurate dating, it is still possible to
give an estimate of the age if the diameter is known and we make some
assumptions on how the SNR evolves with time. Woltjer (1972) developed
a simple model to describe the evolution of a
SNR into a uniform-density ISM. Assuming that an SNR with a diameter of
80 pc is in its {\rm Sedov--Taylor expansion phase}, (in which $t\propto
D^{2.5}$), we have estimated an age of $\sim 10^{5}$ yr for the SNR in
NGC 185.

Type Ib and II SNe are considered to be the final step
of the evolution of massive, rapidly evolving stars,
while type Ia SNe are thought to originate in white dwarfs in binary systems. The rate of the former,
$S_{\rm Ib+II}(t)$, can be calculated from the SFR by 

\begin{equation}
S_{\rm Ib+II}(t)=\int_{m_{\rm SN}}^{\infty}\phi(m)\,\psi_n[t+\tau(m)]\,dm,
\end{equation}

\noindent where $\phi(m)$ is the initial mass function (IMF); $\psi_n(t)$
is the SFR in units of number of stars per year; $\tau(m)$ is the life
time of a star of initial mass, $m$, and $m_{\rm SN}$ is the minimum mass for 
type Ib
and II SN progenitors. As in the remaining of the paper, the criterion is
used that time increases toward the past, the present-day value being
 0, whereas this criterion might seem unnatural,
 it is consistent with the facts that it is easier to work in terms of age, and that we do not know for sure the time at which the galaxy started its evolution. The
SFR in units of number of stars is related to the SFR in mass through
$\psi_n(t)=\psi(t)\times\bar m_\star$, with $m_\star=0.51$ M$_\odot$
being the average stellar mass, obtained from integration of the Kroupa \etal
(1993) IMF.

For NGC 185, the present day Type Ib and II SN rate should be
$S_{\rm Ib+II}(t)=0$ because $\psi(t)=0$ for $t<100$ Myr, which is
much more than the maximum lifetime of the progenitors. In other
words, $\tau(m_{\rm SN})\ll 100$ Myr, which implies that the integral in (4)
vanishes. For this reason, we conclude that the SNR observed in NGC 185
originates in a type Ia event. This is again consistent with the CMD:
the SN having been type Ib or II  would imply a conspicuous MS, even
stronger than that shown in Figure \ref{cm60}.

In any case, it is useful to estimate the average SN rate for
the recent past of the galaxy's history as test of consistency. Let's denote the type Ib+II rate by just $S_{\rm Ib+II}$ and
use the value $\bar\psi_{1-0}=6.6\times 10^{-4}$ M$_\odot$ yr$^{-1}$
obtained in Sec. \ref{sfryoung} for the SFR. Assuming $m_{\rm SN}=10$
M$_\odot$ and using the Kroupa \etal IMF, it turns out that
 $S_{\rm Ib+II}=1.76\times
10^{-3}\bar\psi_{n\,1-0}=2.3\times10^{-6}$ yr$^{-1}$. To this value, the
type Ia SN rate, $S_{\rm Ia}$, must be added to obtain the total rate. It is
estimated that about 50\% of the SNe in spiral galaxies are of type Ia (Tammann 1982). If
this proportion is right for NGC 185 also, we would have $S_{I\rm a}\simeq
2.3\times 10^{-6}$ yr$^{-1}$ and a total SN rate $S_{\rm tot}\simeq
5\times 10^{-6}$ yr$^{-1}$. But it could be larger considering that
NGC 185 is a system with a significant intermediate to old stellar
population. In any case the  values obtained are compatible with only a
SNR having been produced in NGC 185 in the last $10^5$ yr.

\section {The Star Formation History of NGC 185}\label{sfh}

While the characteristics of the recent star formation can be derived in a
relatively simple way from the analysis of the CMD and other photometric
and spectroscopic information, tracing out the SFH for old and even
intermediate-age epochs is by far more complicated. The analysis based
upon synthetic CMDs computed for different input SFHs opens a way to the
determination of the SFH for these older ages. In CMDs such as that used for
 NGC 185 in this paper, old stars are observed in the red tangle, together
with many intermediate-age stars, if present. The bright stars frequently
found over the TRGB in CMDs of dE galaxies may help to discriminate
between the old and intermediate-age SFH. These stars are probably AGBs
with masses over $\sim 1$ M$_\odot$ and are hence intermediate-age, or
possibly even young (younger than $\sim 1$ Gyr).

Bright AGBs have acquired a high relevance as tracers of intermediate-age
stars in dE galaxies ({\it e.g.,} see  Freedman 1992 and Han \etal 1997 for
early and later examples). For this reason we will first discuss to
what extent the presence of such an intermediate-age stellar population
in NGC 185 can be qualitatively deduced from the distribution of bright
AGBs. After that we will make the full analysis of the SFH based
on synthetic CMDs and we will discuss their galactocentric gradients.

\subsection{AGB stars as tracers of the intermediate-age
population}\label{intermediate}

In our $[(V-I),I]$ CMD (see Paper I), a considerable number of bright
stars are observed above the TRGB. These kinds of stars have been observed
optically from the ground in the inner regions of all four Andromeda dE
companions and have several times been interpreted as bright AGBs and as
the signature of an intermediate-age population. Nevertheless, Mart\'\i
nez-Delgado \& Aparicio (1997) have shown that the severe crowding in
the central regions of these galaxies would make the RGB stars  appear
brighter than they are, mimicking the intermediate-age AGB population
\footnote{It must in any case be noted that the absence of the bright
AGB feature in the CMD is not enough to rule out the possibility
of a significant intermediate-age population (see the case for LGS3,
Aparicio {\it et al.} 1997b; or Phoenix, Mart\'\i nez-Delgado, Gallart,
\& Aparicio 1999).}. For this reason, we will restrict the following
analysis to the less crowded regions of the galaxy with $a>177\arcsec$
(where $a$ is the semimajor axis of the concentric ellipses centered on
the galaxy, as defined in Paper I).

In order to shed light on the nature of this population and for
illustrative purposes, we have compared the observed CMD with three
model CMDs computed with constant $\psi(t)$ for different age intervals
(see Sec. \ref{metodo}). Observational effects for the $177\arcsec<a<236
\arcsec$ have been simulated using the procedure explained in Paper
I. Figure \ref{compa} displays the observed CMD for this region and the
model CMDs for the different age intervals.

 The explicit and realistic simulation of the observational
effects performed in these model CMDs implies that we can directly compare
their visual aspect, as well as the star counts, to those of the observed
CMD. A simple visual inspection indicates that the model CMD for the 15--9 
Gyr old population (Fig. \ref{compa}a) cannot by itself alone account
for the presence of the stars brighter than $I\sim 20.5$ mag but that some intermediate-age population is needed. Since some of
those stars could also be foreground stars, we have counted the stars
in a region situated above the TRGB with $19$ mag $<I<$ 20 mag and 
$1.2<(V-I)<4.2$
in the CMD of NGC 185 and in a companion field. The results for several
elliptical annuli of NGC 185 ($N_{\rm AGB}$) and for the comparison field
($N_{\rm f}$) after scaling to the corresponding areas, are given in
Table 3. The errors listed in this table have been calculated assuming
Poissonian statistics. Table 3 shows that the foreground contribution
is not enough to account for the number of stars observed in the CMD
of NGC 185 in any of the $a >177 \arcsec$ annuli, supporting the idea
that an intermediate-age population is required. This is not in
contradiction with the result by Geisler {\it et al.} (1999), who report
{\it HST} observations of an inner and an outer region of NGC 185 in which
upper-AGB stars are absent (see Sec. \ref{result}). This result could
be due to the small field of the {\it HST} Planetary Camera ($36.8\times 36.8
(\arcsec)^2$). Indeed, using this and our results listed in Table 3,
only four stars are expected above the TRGB in the interval $19$ mag $<I<$ 20
mag,
$1.2<(V-I)<4.2$ for the outer field of Geisler \etal

\subsection {Method and hypotheses}\label{metodo}

The detailed derivation of the SFH from synthetic CMDs requires
deeper data than those we have available for NGC 185 (see Gallart \etal
1999). However an estimate of the old, intermediate-age and young SFR is
still possible from our CMD of the galaxy. We have followed the method
described by Aparicio \etal (1997b). In short, the SFH is considered to
be composed of three functions: the SFR, $\psi(t)$, the IMF, $\phi(m)$, and
the chemical enrichment law, $Z(t)$. A function $\beta(f,q)$ accounting for
the fraction and mass distribution of binary stars can also be considered
(see Gallart \etal 1999). The functions
$\phi(m)$, $Z(t)$ and $\beta(f,q)$ are treated
as input information to the models. Several combinations of them can be
checked. For each of them, the best solution or a group of good solutions
are obtained for $\psi(t)$, in the following way. First, several simple
{\it partial} models are obtained for the combination of $\phi(m)$,
$Z(t)$ and $\beta(f,q)$. Each partial model contains stars in a given,
relatively narrow age interval with constant SFR. {\it Global} models can
then be computed by linear combinations of the partial models in the form
$\psi(t)=l\sum a_i\psi_i$, where $\psi_i$ is the SFR of partial model $i$,
$a_i$ is the linear combination coefficient and $l$ is a normalization
constant. Secondly, several regions are defined in the CMD sampling areas
populated by stars of different ages and in different stellar evolution
phases, and the numbers of stars in these regions in the observational
CMD and in each of the partial models are counted. The star counts
corresponding to any of the global model can be then computed and compared
with those of the observational CMD through a merit function. Finally,
the best or a group of good global models is found out by minimization of
the merit function with respect to the $a_i$ coefficients.
This procedure gives several solutions for the SFH producing model CMDs
compatible with the observational one. The solution is not unique, but
the range covered by them indicates the uncertainty of the result and
allows the elimination of impossible scenarios for the SFH.

In the particular case of NGC 185, data are not deep enough to seek a
detailed description of the SFH. For this reason we intend to find out
only estimates of the old, intermediate and young SFH and therefore,
we have simplified the number of inputs for $\phi(m)$, $Z(t)$ and
$\beta(f,q)$.  In particular, we have used only one shape for $\phi(m)$:
the one by Kroupa {\it et al.} (1993) and we have neglected the effects
of binary stars. The later is justified by the fact that we are dealing
mainly with evolved RGB and AGB stars. For them, only binaries with
$q\simeq 1$ can noticeably affect the distribution of stars in the
CMD. This is because, in other case, binaries in which the least massive
component is in the RGB or AGB have likely a massive component having
evolved into the white dwarfs phase, while binaries with the most massive
component in the RGB or AGB have probably a secondary in the low MS,
thereby very weakly affecting the total luminosity of the system.

 In spite of the age--metallicity degeneracy in the RGB phase, limits
can be set on $Z(t)$ if two simple assumptions are made: that no stars
can exist in the galaxy with metallicities implying a red tangle different
(in shape, wideness or position) from the observed one, and that $Z(t)$ is
not a decreasing function of time.  With these assumptions,
three {\it partial} model CMDs with the same constant SFR were computed
for three age intervals (in Gyr): $1\leq t<3$, $3\leq t<9$, and $t\geq
9$, (see for example Fig. \ref{compa}). Stars in each partial model have
been computed with metallicities taken at random inside an interval such
that the position and the full width of the red tangle are matched. In
this way, independently of the age distribution resulting for global
model, the restriction that the position and shape of the red tangle
must be well reproduced will be satisfied. Of course,
other choices of $Z(t)$ could also (but not always) produce good results
if combined with adequate $\psi(t)$. However, the low time resolution
that we can achieve implies that the task of looking for those solutions
is not justified. Since low time resolution means averaging $\psi(t)$
for long intervals of time, we believe that the solution obtained here
is a good approximation to reality.

The comparison between model and observed CMDs is made through star
counts in the three selected regions defined in Figure \ref{zonas}. These
regions were chosen to sample those stellar evolutionary phases that are
sensitive to the distribution of stellar ages that can be most efficiently
discriminated with our data. In particular, region 1 would be populated
by old and intermediate-age RGB and AGB stars; region 2 would contain
old and intermediate-age AGB stars; and region 3 would be populated by
bright, intermediate-age AGB stars only.

\subsection {The resulting SFH}\label{result}

As discussed in Paper I, the CMD of the inner regions of NGC 185 is
strongly affected by crowding. For this reason, we have divided our study
in two parts. For the inner $177\arcsec$ (semimajor axis; see Paper I for
the definition of several annular elliptical regions in NGC 185), a simple
estimate of the total averaged $\psi(t)$ has been obtained from the number
of stars in the upper part of the red tangle (Aparicio, Bertelli, \& Chiosi 1999) . This estimate has been made from two annular
regions: i) inner-A ($a<118\arcsec$) and ii) inner-B ($118\arcsec\leq
a<177\arcsec$). The resulting SFRs are $\bar{\psi}_{\rm tot}=8.2\times 10^{-3}$
M$_{\sun}$ yr$^{-1}$ or $\bar{\psi}_{\rm tot}/A=3.3\times 10^{-8}$ M$_{\sun}$
yr$^{-1}$ pc$^{-2}$ for region inner-A and $\bar{\psi}_{\rm tot}=4.6\times
10^{-3}$ M$_{\sun}$ yr$^{-1}$ or $\bar{\psi}_{\rm tot}/A=1.9\times 10^{-8}$ 
M$_{\sun}$ yr$^{-1}$ pc$^{-2}$ for region inner-B.

For the outer ($177\arcsec\leq a$) regions the analysis of the SFH has
been done as described in Sec. \ref{metodo} for three elliptical annular
regions defined as follows (see Paper I): i) outer-A ($177\arcsec\leq
a<236\arcsec$); ii) outer-B ($236\arcsec\leq a<295\arcsec$); and iii) outer-C,
($295\arcsec\leq a$). The latter includes the two outer annuli defined
in Paper I.

The observed CMDs of the outer regions are displayed in
Figure \ref{3cmd}. In these CMDs, a number of blue stars (bluer that
$(V-I)\sim 1.4$) and brighter than $I\sim 22$ mag are present. From estimates
made from the foreground field CMD shown in Fig. \ref{cmdcomp} (see
Sec. \ref{resolved}), these stars are very probably foreground stars and
will not be further considered in SFH analysis.

The main result arising from our analysis is that the bulk of the stellar
population of NGC 185 formed at an early epoch, but that star formation
has continued at a lower rate for almost the rest of the lifetime of
the galaxy. Figure \ref{sfr} shows the adopted solutions for $\psi(t)$.
Each of them has been obtained averaging the global models matching the
stellar counts of the observational CMD to better than 1.5 $\sigma$. The
results are also listed in Table 4, together with the estimates for
the inner regions. The SFR for young stars in the innermost region obtained
in Sec. \ref{sfryoung} has been also included. The SFR values for
each age interval listed in the first four lines have been used to
obtain the averaged SFRs for the complete period of star formation
($\bar{\psi}_{15-1}$) given in line 5. Lines 6 to 10 give the same
as lines 1 to 5, after normalization to the area of the region in
pc$^2$.

Our results for the SFH  can be compared with deeper {\it HST}
observations. Geisler {\it et al.} (1999) present preliminary results
for the field-star CMDs for two regions in the vicinity of the NGC 185
globular clusters FJJIII ({\it inner} field) and FJJIV/V ({\it outer}
field). Both diagrams show a striking red clump (RC) and a small number
of blue horizontal branch (HB) stars. The outer field of Geisler \etal
falls
inside our outer region-A. We have explicitly computed the model CMD resulting
from the SFH we have obtained for this region, but simulating it to be
$\sim 3$ mag deeper than our observational CMD. The resulting model
CMD is shown in Figure \ref{HST} and represents what would be obtained from
deep {\it HST} observations if our results for the SFH are correct. The visual,
qualitative agreement with the {\it HST} CMD of the FJJIV/V field (Geisler,
private communication) is very good, especially in the HB and RC morphology
 but also regarding the width and overall structure of the
RGB. It must be noted that the RC [$(V-I)\sim 1.1$; $I\sim 24.0$] of the
model CMD shown in Fig. \ref{HST} is mainly populated by intermediate-age
stars. Consequently, the RC obtained from {\it HST} data is a good alternative
observational confirmation of the star formation having taken place at
intermediate ages in NGC 185, as  has been found in our analysis.

\subsection{Galactocentric gradients of the SFR}\label{grad}

The values given for $\bar{\psi}_{\rm tot}/A$ in the last line of Table
4 indicate a gradient of decreasing SFR with galactocentric radius. We
have studied how this variation is correlated with the surface brightness
(SB) profile of NGC 185, in order to search for possible gradients of
the mass-to-light ratio within the galaxy. These would in turn indicate
gradients in the composition of the stellar population, and hence in the
shape of the SFH.

Figure \ref{sb} shows the variation of $\log(\bar{\psi}_{\rm tot}/A)$ as
a function of galactocentric radius (filled dots) together with the
same for the mean SB $\mu_{r}$ (open circles) obtained after averaging
the $\mu_{r}$ profile by Kent (1987) for each annulus. Both quantities
follow a similar trend which can be interpreted as the relative amounts
of stars of different ages being similar for each region. This is in good
agreement with the relatively small changes in the shape of $\psi(t)$
obtained for the outer regions, and the fact that it can reasonably be
extended inward indicates that,  overall,  shape would have been
similar in the inner regions too. This could be in apparent contradiction
with the young population having been found close to the center of the
galaxy only, but this may be compensated by the fact that the total
life-averaged $\psi(t)$ is also bigger for the innermost region.

\section{Young stars and the origin of the gas in NGC 185}\label{gas}

The origin of the raw gas from which the young stars are born in E and dE
galaxies is a puzzling question. Although E galaxies were long thought to
be devoid of gas, Faber \& Gallagher (1976) showed that the mass returned
during the galaxies' lifetimes to the interstellar medium by evolved,
dying stars should be significant. However, it is found that the H~{\sc i} in
E galaxies is highly extended with respect to the optical components and
is rotating with angular momenta per unit mass much higher than that of
the stars. This points to an external origin for the gas in these systems
(van Gorkom 1992).

The case may be different for dE. Young \& Lo (1997) find different
scenarios for the three dEs they analyze: no gas in NGC 147; gas having
different angular momentum per unit mass from stars in NGC 205 and gas
kinematically compatible with that of the stellar component in
NGC 185. For the latter, in particular, they find a velocity dispersion
$\sigma_v=15.3$ km s$^{-1}$ for the gas, somewhat smaller than the stellar
one: $\sigma_v=22$ km s$^{-1}$ (Bender \etal 1991). The results for the
SFH we have obtained in the previous sections can shed light on the
question of the external or internal origin of the gas in NGC 185.

The rate at which the material is returned to the ISM by dying stars
in the central region of NGC 185 can be evaluated from integration of
the SFR through $$R(t)=\int_{m_{\rm l}}^{m_{\rm u}} [m-p(m)]\phi(m)\psi[t+\tau(m)]dm$$,

\noindent where $m_{\rm l}$ and $m_{\rm u}$ are lower an upper limits,
respectively, for the stellar
mass and $p(m)$ is the mass of the remnant of a star of initial mass
$m$. Note that within this paper the criterion is used that time increases 
toward
 the past and the present day value is 0. Using $\psi(t)=8.2\times 10^{-3}$ M$_\odot$ yr$^{-1}$ for 1 Gyr $\leq
t<15$ Gyr and $\psi(t)=6.6\times 10^{-4}$ M$_\odot$ yr$^{-1}$ for 0 Gyr $\leq
t\leq 1$ Gyr, and the relations given by Vassiliadis \& Wood (1993) for $p(m)$, we
obtain for the present time $R(0)=10^{-3}$ M$_\odot$ yr$^{-1}$, which is
1.5 times the recent, averaged SFR $\bar\psi_{1-0}$. It is worth noting
that no significant differences are found if the integration is stopped
at $t=0.1$ Gyr, and that almost 90\% of the returned material comes from
stars born in the 15 Gyr $\leq t<1$ Gyr interval.

Making a detailed kinematical evolutionary model is beyond the scope of
this paper, but some other pieces of information can be qualitatively
evaluated. The gas content observed in the central region of NGC 185
($7.3\times 10^5$ M$_\odot$) can be built up in about 2 Gyr from the
ejected gas not used for star formation. But for this picture to work
it is also necessary that SN explosions and high-velocity winds from
massive stars are neither efficient enough to sweep away the ISM nor to
heat it up over shorter time-scales. Lozinskaya (1992) shows that one SN
of energy $10^{51}$ erg would transfer some $5\times 10^{49}$ erg to
the interstellar medium. But this energy represents only 10\% of the
kinetic energy of the gas in NGC 185 (Young \& Lo 1997). As we have
shown, this amount of energy would be injected into the ISM at a rate
of once every $5\times 10^5$ yr. On the other hand, Chevalier (1974)
showed that the cooling time-scale for typical SNR is of the order of
$10^5$ to $10^6$ yr, likely short enough to allow NGC 185 maintaining
the cool gas component it shows.

Summarizing, we have a picture in which the kinematical properties of
gas and stars are consistent with the internal origin of the former. The
rate at which evolved stars return gas to the ISM is enough to seed the
recent star formation observed in the center of the galaxy and the SN
rate is probably low enough to allow the galaxy to retain the gas not
used in the new stellar generations.

Another puzzling question is why NGC 147, a galaxy that could be
considered similar to NGC 185, with a large amount of old, dying stars,
shows traces of neither gas nor recent star formation (see Young \& Lo
1997 for the gas and Han \etal 1997 for the stellar population). Resolving
this question will require a more detailed analysis of the properties of
NGC 147 and could perhaps indicate that mechanisms such as gas removing by SN explosions and subsequent delay of star formation would be at work (see 
Burkert \& Ruiz-Lapuente 1997).

\section{Global properties of NGC 185} \label{global}

The SFR $\psi(t)$ can be used together with the distance obtained in
Paper I and data from other authors, to calculate integrated properties of
NGC 185. Table 5 gives a summary of these properties. Bracketed figures
refer to the bibliographic sources for each data. The first two lines list values for $\psi(t)$
averaged for several intervals of time: $\bar\psi_{15-0}$ is the average
for the whole life of the galaxy; $\bar\psi_{1-0}$ corresponds to the
last 1 Gyr. As the field covered by our observations is smaller than
the galaxy size ($R_{t}=16\arcmin$, Hodge 1963), the resulting SFR has
been multiplied by a scale factor obtained as the rate between the total
luminosity of the galaxy (Sandage \& Tammann 1981) and the luminosity
covered by our field estimated from the surface-brightness profile given
by Kent (1987). Lines 3 and 4 give the same values normalized to the
area. Line 5 gives the $\bar\psi_{1-0}/A$  value obtained for the central
region in which the recent star formation is detected,  in which area
is $A_{\rm inner}=150$ pc $\times 90$ pc (see Sec. \ref{resolved}).

The distance to the Milky Way and to the barycenter of the Local
Group are given afterwards, followed by the absolute magnitudes and
the total luminosities in different bands, calculated for the adopted
distance. $M_\star$ is the mass in stars and stellar remnants, calculated
from integration of $\psi(t)$ and assuming that a fraction 0.8 of this integral
remains locked into stellar objects. To obtain $M_{\rm gas}$, the H~{\rm i}
mass has been multiplied by 4/3 to account for the He content. $\mu$ is
the gas fraction relative to the total mass intervening in the chemical
evolution. Finally, the gas and total mass to luminosity fractions
are given.

\section {Conclusions}\label{conclusion}

The SFH and the properties of the dE galaxy NGC 185 have been analyzed in
the light of new ground-based $H_{\alpha}$ and $BVI$ photometry. The SFH
has been estimated for old, intermediate-age, and young stars and its spatial
variations investigated using synthetic CMDs, detailed simulation of
observational effects and reliable estimates of the foreground stellar
contamination.

The recent star formation is confined into the central $150 \times 90$
pc$^2$ of the galaxy. After correction for foreground star 
contamination, the
$[(B-V), V]$ CMD of the central field ($3.75 \arcmin \times 3.75\arcmin$)
shows a population of blue/yellow stars that indicates star formation
in the last 1 Gyr. The youngest stars were probably  formed some
100 to 125 Myr ago and the SFR since 1 Gyr ago is estimated to be
$\bar{\psi}_{1-0}=6.6\times 10^{-4}$  M$_{\sun}$ yr$^{-1}$.

The luminous blue stars discovered by Baade (1951) in the center of NGC
185 have been discussed using new $B$ CCD images and Baade's  original
photographic plates. A total of 19 blue, bright objects have been found
from a visual inspection of these plates in the central $20 \arcsec$ of
the galaxy. We believe that most of these objects are the blue, bright
stars originally mentioned by Baade. We have ruled out the possibility
that they are foreground stars. They lie in a region where evolved, core
He-burning, 40 to 150 Myr old stars are expected, but, if this were their
age range, they should be accompanied by a conspicuous MS, which is not
present. Based on this and on the fact that several of them show a fuzzy,
unresolved aspect, with an FWHM larger than the stellar one, we conclude
that most of these objects are young stellar clusters. In fact, at the
distance of NGC 185, their linear sizes, color and absolute magnitudes
are consistent with those of typical young Galactic open clusters.

We have presented new $H_{\alpha}$ images of the central region of the
galaxy showing the SNR discovered by Gallagher {\it et al.} (1984) with
more detail.  It has an arclike morphology and might be a portion of a
larger, old remnant  80 pc in diameter. We estimate its $H_{\alpha}$
luminosity to be 1.3 $\times 10^{36}$ erg s$^{-1}$ and its age $\sim
10^{5}$ yr. Assuming that it proceeded from an SN~II, we estimate the SFR
in the central region to be in the range $3\times 10^{-3}$ to $6\times
10^{-3}$  M$_{\sun}$ yr$^{-1}$, almost one order of magnitude larger than
the recent SFR estimated from the resolved young population. Again, this
would likely require a conspicuous MS to be present in the CMD and hence
compels us to conclude that the SNR comes from an SN~Ia progenitor. 

We have analyzed the SFH for two inner and three outer regions in NGC
185 using synthetic CMDs and an empirical simulation of observational
effects using the results of a large number of artificial-star tests. The
resulting scenario shows that the bulk of the stars in NGC 185 were
formed in an early epoch. The star formation has continued at a much
lower rate for intermediate ages. This produces a population of stars above
the TRGB. They are relatively bright AGBs and cannot be accounted for by
foreground contamination. No traces of star formation more recent than 1
Gyr are found outside the central region of the galaxy, where the star
formation  continued until some 100 Myr ago. The absence of younger
stars could be interpreted as the star formation proceeding in small,
mild bursts, taking place with intervals of several 100 Myr between them.

New stars are produced only in the very central region of the galaxy. This
could indicate that young stars are born from processed material ejected
by old, dying stars. Calculations show that the rate at
which evolved stars return gas to the ISM is enough to seed the recent
star formation observed in the center of the galaxy and the SN rate
is probably low enough to allow the galaxy to retain the gas not used
in the new stellar generations. This, together with the fact that the
kinematical properties of gas and stars are similar, supports the idea of
an internal origin for the gas component observed in the central regions
of the galaxy.

 Finally, we have derive several
integrated global parameters of NGC 185 which are listed in Table 5.

\acknowledgments

We are indebted  to Dr. Allan Sandage for useful indications of which objects probably Dr. Baade referred to
as being {\it the blue stars} in the Baade's original plates of
NGC~185. We thank the enjoyable and 
interesting discussions, and his patience teaching us how to look at the
plates, and making possible to obtain some useful reproductions of them. 
The Observatories of the Carnegie Institution of Washington are also
acknowledged for making available to us  Baade's original plates of
NGC 185 taken with the 5 m Hale Telescope at Mount Palomar, California
(USA).  We also thank to the anonymous referee for careful reading of the original manuscript and for useful comments and suggestions. We are indebted 
to Dr. Geisler for useful comments and for making
available to us his preliminary {\it HST}
 results on NGC 185. This work has been
substantially improved after fruitful discussions with Drs. Bertelli,
Chiosi, Da Costa, Gordon, Freedman, Ho, Welch and Young to whom we are
very grateful. AA thanks the Observatories of the Carnegie Institution
and its staff at Pasadena for their hospitality during his one year
term stay as a visiting researcher. During that period, AA was funded
by the Direcci\'on General de Ense\~nanza Superior, Investigaci\'on y
Desarrollo of the Kingdom of Spain (Grant PB95-554). DMD also thanks the
staff of the Observatories of the Carnegie Institution at Pasadena for
their hospitality during his visit. This work is part of the PhD thesis
of DMD and has been financially supported by the Instituto de Astrof\'\i
sica de Canarias (Grant P3/94) and the Direcci\'on General de Ense\~nanza
Superior, Investigaci\'on y Desarrollo of the Kingdom of Spain (Grant
PB94-0433).

\newpage

\figcaption[figure1.ps]{Residuals of the transformation to
standard $B$ magnitudes using the photometry of Lee {\it et al.}
(1993).\label{calibraB}}

\figcaption[figure2.ps]{Color image of the central region of NGC 185
obtained from a combination of $B$, $V$ and $R$ images. The size of the field is $2.8\arcmin \times 2.8\arcmin$. North is up and east is to the left. \label{color}}

\figcaption[figure3.ps]{$[(B-V), V]$ CMD of the central ($3.75 \arcmin
\times 3.75\arcmin$) region of NGC 185.\label{cmd}}

\figcaption[figure4.ps]{$[(B-V), V]$ CMD of a comparison field situated
25$\arcmin$ north of the center of NGC185. \label{cmdcomp}}

\figcaption[figure5.ps]{Crowding-corrected $V$ luminosity function for
blue--yellow stars in NGC 185 (solid line) and in the comparison foreground
field (dotted line). \label{lf}}

\figcaption[figure6.ps]{Synthetic ({\it panel A}) and model ({\it panel B})
 CMDs
computed with constant star formation rate $\psi(t)$ for the interval of time from 15 to
0.125 Gyr ago. Stellar metallicities take random values in the interval
$0.001\leq Z\leq 0.01$. \label{cm125}}

\figcaption[figureX.ps]{ $[(B-V), V]$ CMD of the central ($3.75 \arcmin
\times 3.75\arcmin$) region of NGC 185. An isochrone from the Padua
library (Bertelli {\it et al.} 1994) for Z=0.001 and age 125 Myr is
overplotted. Baade's blue {\it stars} are displayed as large
dots, with pentagon shaped ones being those showing an unresolved
appearance. The position in the CMD of the Pleiades
(open triangle), Praesepe (open square), and NGC 188 (open circle) 
 Galactic open clusters at
the distance and reddening of NGC 185 are also plotted. \label{cmd_blue}}

\figcaption[figure7.ps]{Magnification of the central part of photographic
plate number PH662B of NGC 185 taken by W. Baade in 1952 August
20/21 with the Palomar 200$''$ telescope. A 103a0 emulsion was used with a
GG1 filter. The exposure time was 30 minutes. North is up and east is to the left. The size of the field is $3.2 \arcmin
\times 3.2\arcmin$. Baade wrote in the plate cover:
{\it Incipient resolution in photographic light clearly visible}. This
plate is published for the first time here. \label{placa}}

\figcaption[figure8.ps]{Finding chart for the 19 blue stars selected
from the original Baade plates in the central region of NGC 185 listed in Table 2. The
seeing is $0.6\arcsec$. North is up and east is to the left. The size of the field is $2.5 \arcmin \times 2.5\arcmin$.\label{map}}

\figcaption[figure9.ps]{Synthetic ({\it panel A}) and model ({\it panel B}) CMD,
computed with constant $\psi(t)$ for the interval of from 1 to 0.04 Gyr
ago. Stellar metallicities take random values in the interval $0.001\leq
Z\leq 0.01$. \label{cm60}}

\figcaption[figure10.ps]{Continuum-subtracted $H_{\alpha}$ image of the
center of NGC 185. The exposure time is 900 s. North is up and east is
to the left. The size of the field is $2.3\arcmin
\times 2.3\arcmin$. The cross marks approximately the center of the galaxy. The two compact objects close to the center are possible
previously unknown planetary nebulae. The three similar-looking objects
visible at the periphery of the figure are planetary nebulae catalogued
by Ford {\it et al.} (1973). \label{halfa}}

\figcaption[figure11.ps]{Comparison of the observed CMD for the annulus
$177\arcsec\leq a<236\arcsec$ (upper panel) with partial model CMDs for three age
intervals (lower panels). The figure suggests that the bright red stars above the TRGB
of the observational diagram are probably intermediate-age AGB stars, 
which is confirmed by subsequent investigation of the SFH. \label{compa}}

\figcaption[figure12.ps]{The regions used for the determination of the
SFH overplotted onto the CMD of the outer regions ($a>177\arcsec$)
of NGC 185.\label{zonas}}

\figcaption[figure13.ps]{Observed $[(V-I), I]$ CMDs for the three regions
in which the outer part of NGC 185 has been divided for the analysis of
the SFH (see text for details). \label{3cmd}}

\figcaption[figure14.ps]{SFRs of regions outer-A, outer-B and outer-C
as a function of time.  \label {sfr}}

\figcaption[figure16.ps]{Model CMD produced by the SFH shown in Fig. 15
({\it top panel}) for a limited magnitude about 3 mag deeper than that reached
by our photometry. This is approximately what would be seen from deep
{\it HST} observations if our results for the SFH are correct. \label{HST}}

\figcaption[figure17.ps]{Radial variation for the $\log(\bar{\psi}/A)$
compared with the SB profile of NGC 185 obtained by Kent (1987) averaged
for each annulus.\label{sb}}

\newpage

\begin{deluxetable}{cccccc}
\tablenum{1}
\tablewidth{0pt}
\tablecaption{ Journal of observations}
\tablehead{
\colhead{Date}      & \colhead{Object} &
\colhead{Time(UT)}          & \colhead{Filter}  &
\colhead{Exp. time (s)} &        
\colhead{ FWHM ($\arcsec$)}}
\startdata
97.07.27 & NGC 185 & 03:43 & $H_{\alpha}$-cont & 900 & 0.7 \nl
97.07.27 & NGC 185 & 03:59 & $H_{\alpha}$ & 900 & 0.7 \nl
97.07.27 & NGC 185 & 04:14 & $H_{\alpha}$ & 900 & 0.7 \nl
97.07.29 & NGC 185 & 04:37 & $B$ & 900 & 0.6 \nl
97.07.29 & FIELD & 04:55 & $I$ & 400 & 0.5 \nl
97.07.29 & FIELD & 05:02 & $V$ & 600 & 0.5 \nl
97.07.29 & FIELD & 05:13 & $B$ & 900 & 0.6 \nl
\enddata
\end{deluxetable}

\newpage

\begin{deluxetable}{ccccccccccc}
\tablenum{2}
\tablewidth{0pt}
\tablecaption{Baade's blue stars in NGC 185}
\tablehead{
\colhead{$N$}   & \colhead{$N_{Lee}$}   &
\colhead{$X$}  & \colhead{$Y$}          & 
\colhead{$B$} & \colhead{$V$}          & 
\colhead{$I$} &  \colhead{$(B-V)$} & 
\colhead{$(V-I)$} &  \colhead{FWHM ($\arcsec$)} & Note}
\startdata
1 & - & 108.37 & 346.03 & 21.56 & 21.14 & 20.55 & 0.42 & 0.59 &  0.7 &\nl
2 &  3802 &171.98 & 291.21 &  21.63 & 20.96 &  19.37& 0.67 & 1.59 & 1.4 & \tablenotemark{a} \nl
3  &  - &220.05 & 283.78 &   19.90 &    19.55 & - &  0.35& - & 1.8 & \tablenotemark{a}\nl
4 & 3116 & 221.23 & 260.47 &    20.20 &  19.79 & - & 0.41  & - & 0.7 &\nl     
5 & 2244 & 229.66 & 215.15 &  21.70 & 21.01 &   20.07 & 0.69 & 0.94 & 1.1 & \tablenotemark{a} \nl
6 & 2468 & 245.62 & 228.42 &   21.40 &    20.92 &   19.18 & 0.48 & 1.74 & 1.2 & \tablenotemark{a}\nl
7 & 2452 &  251.97 & 228.17 & 21.80 &  20.67 & 19.39  &  1.13 & 1.28 & - &\nl
8 & 2588 & 251.58 & 235.00 &   21.78 &    21.48 &    - & 0.30 & - & - & \tablenotemark{a}\nl
9 & - & 243.85 &254.91 &   20.80 &  20.00 & - & 0.80 & - & - & \tablenotemark{a}\nl
 10 & - & 253.21 & 248.88 &    20.98 &   20.90 &   - & 0.08 & - & - & \tablenotemark{b} \nl
 11 & 2933 & 252.93 & 252.16 &   21.01 &  20.22 &   19.30 & 0.79 & 0.92 & - & \tablenotemark{a}\nl
 12 & 1839 & 276.22 & 193.54 &   21.25 &    20.66 &   19.62& 0.59 & 1.04 & 2.5& \tablenotemark{a}\nl
 13 & - & 263.20 & 242.87 &   21.92 &    20.89 &  19.32 &1.03 & 1.57 & - & \tablenotemark{a} \tablenotemark{b} \nl        
 14 & 5370 & 277.79 & 380.95 &   21.26 &    20.36 &   19.58 & 0.90 & 0.78 & 0.7 &\tablenotemark{c} \nl    
 15 & - & 273.01 & 264.95 &   21.70 &    20.72 &   19.33 & 0.98& 1.39 & -& \tablenotemark{a} \tablenotemark{b}\nl
 16 & 3095 & 285.02 & 260.89 &   21.30 &    21.24 &   20.64 & 0.06 & 0.60 & - &\tablenotemark{a} \tablenotemark{b} \nl
 17 & 4084 & 306.23 & 309.79 &   21.28 &    20.78 &   19.86 & 0.50 & 0.92 & 1.2 & \tablenotemark{a} \nl
 18 & 1369 & 373.77 & 169.53 &   20.78 &    20.46 &   20.12 &0.32  & 0.34 & 0.6 &\nl
 19 & 5683 & 325.68 & 406.92 &   21.69 &    21.25 &   20.92 & 0.44 & 0.33 & 0.6 &\nl
\enddata
\tablenotetext{a}{Fuzzy, unresolved appearance.}
\tablenotetext{b}{Immersed in a dust cloud, possibly with high reddening.}
\tablenotetext{c}{Planetary nebula. }

\end{deluxetable}

\newpage

\begin{deluxetable}{ccc}
\tablenum{3}
\tablewidth{0pt}
\tablecaption{AGB stars and contamination by foreground stars}
\tablehead{
\colhead{ } &    
\colhead{$N_{\rm AGB}$} & 
\colhead{$N_{\rm f}$}}

\startdata
$ 177\arcsec < a < 236\arcsec$  & 28 $\pm$ 5 & 7 $\pm$ 3 \nl
$ 236\arcsec < a < 295\arcsec $ & 27 $\pm$ 5 & 7 $\pm$ 2 \nl
$ 295\arcsec < a < 354\arcsec $ & 15 $\pm$ 4 & 5 $\pm$ 2 \nl
$a>354\arcsec$             & 11 $\pm$ 3 & 5 $\pm$ 2  \nl
\enddata
\end{deluxetable}

\begin{deluxetable}{lccccc}
\tablenum{4}
\tablewidth{0pt}
\tablecaption{Star formation rates }
\tablehead{\colhead{ } & \colhead{Inner A } &
\colhead{Inner B } &
\colhead{Outer A} &
\colhead{Outer B} &
\colhead{Outer C}}

\startdata

$\bar{\psi}_{15-9}$ ($ 10^{-4}\ {\rm  M}_{\sun} {\rm yr}^{-1}$) & - & - &  23.3 &  13.2 & 16.0 \nl
$\bar{\psi}_{9-3}$ ($ 10^{-4} \ {\rm M}_{\sun} {\rm yr}^{-1}$)& - & - & 1.0 & 3.7 &  1.7 \nl
$\bar{\psi}_{3-1}$  ($ 10^{-4}\ {\rm M}_{\sun} {\rm yr}^{-1}$)& - & - & 3.7 & 1.5 & 0.5\nl
$\bar{\psi}_{1-0}$  ($ 10^{-4}\ {\rm M}_{\sun} {\rm yr}^{-1}$) & 6.6 & -& - & -  & -  \nl
$\bar{\psi}_{tot}$ ($ 10^{-4} \ {\rm M}_{\sun} {\rm yr}^{-1}$) & 82 & 46 & 11.0 & 7.5 & 7.7\nl
$\bar{\psi}_{15-9}$ /A ($ 10^{-9} \ {\rm M}_{\sun} {\rm yr}^{-1} {\rm pc}^{-2}$) & - & -  & 8.7 &  4.4 &   3.2  \nl
$\bar{\psi}_{9-3}$ /A ($ 10^{-9} \ {\rm M}_{\sun} {\rm yr}^{-1} {\rm pc}^{-2}$) & - & - & 0.4  & 1.2 & 0.3 \nl
$\bar{\psi}_{3-1}$ /A ($ 10^{-9} \ {\rm M}_{\sun} {\rm yr}^{-1} {\rm pc}^{-2}$)  & - & - & 1.4 & 0.5 & 0.1  \nl
$\bar{\psi}_{1-0}$ /A ($ 10^{-9} \ {\rm M}_{\sun} {\rm yr}^{-1} {\rm pc}^{-2}$)  & 2.6 & - & - & - & -   \nl
$\bar{\psi}_{tot}$/A ($ 10^{-9} \ {\rm M}_{\sun} {\rm yr}^{-1} {\rm pc}^{-2}$)& 33 & 19 &  4.1 & 2.5 & 1.5 \nl
\enddata
\end{deluxetable}

\newpage

\begin{deluxetable}{llr}
\tablenum{5}
\tablewidth{0pt}
\tablecaption{Global properties of NGC 185
\label{global}}
\tablehead{
\colhead{~} &  \colhead{~} & \colhead{NGC 185}}
\startdata
$\bar\psi_{15-0}$ & ($10^{-4}$ M$_\odot$yr$^{-1}$) & 247 (1) \nl
$\bar\psi_{1-0}$ & ($10^{-4}$ M$_\odot$yr$^{-1}$) & 6.6 (1) \nl
$\bar\psi_{15-0}/A$ & ($10^{-9}$ M$_\odot$yr$^{-1}$pc$^{-2}$) & 1.22
(1)  \nl
$\bar\psi_{1-0}/A$ & ($10^{-9}$ M$_\odot$yr$^{-1}$pc$^{-2}$) & 0.03
(1)  \nl
$\bar\psi_{1-0}/A_{\rm inner}$ & ($10^{-9}$ M$_\odot$yr$^{-1}$pc$^{-2}$) & 49
(1)  \nl
$d$ & (Mpc) & 0.616 (1)  \nl
$d$ (LG) & (Mpc) & 0.178 (1) \nl
$M_{B}$ & & --14.63 (2)  \nl
$M_{V}$ & &  --15.52 (3) \nl
$L_B$ & ($10^8$ L$_\odot$) & 1.40 \nl
$L_V$ & ($10^8$ L$_\odot$) & 1.25  \nl
$M_\star$ & ($10^8$ M$_\odot$) & 2.96 (1) \nl
$M_{\rm gas}$ & ($10^8$ M$_\odot$) & 0.0073 (4,5)  \nl
$\mu=M_{\rm gas}/(M_\star+M_{\rm gas})$ & & 0.025 (1) \nl
$M_{\rm gas}/L_V$ & (M$_\odot$/L$_\odot$) & 0.006  \nl
\enddata
\tablerefs{(1) This paper; (2) Sandage \& Tammann 1987; (3) Kodaira {\it et al.} 1987;
 (4) Young \& Lo 1997; (5) Sage, Welch \& Mitchell 1998 }
\end{deluxetable}

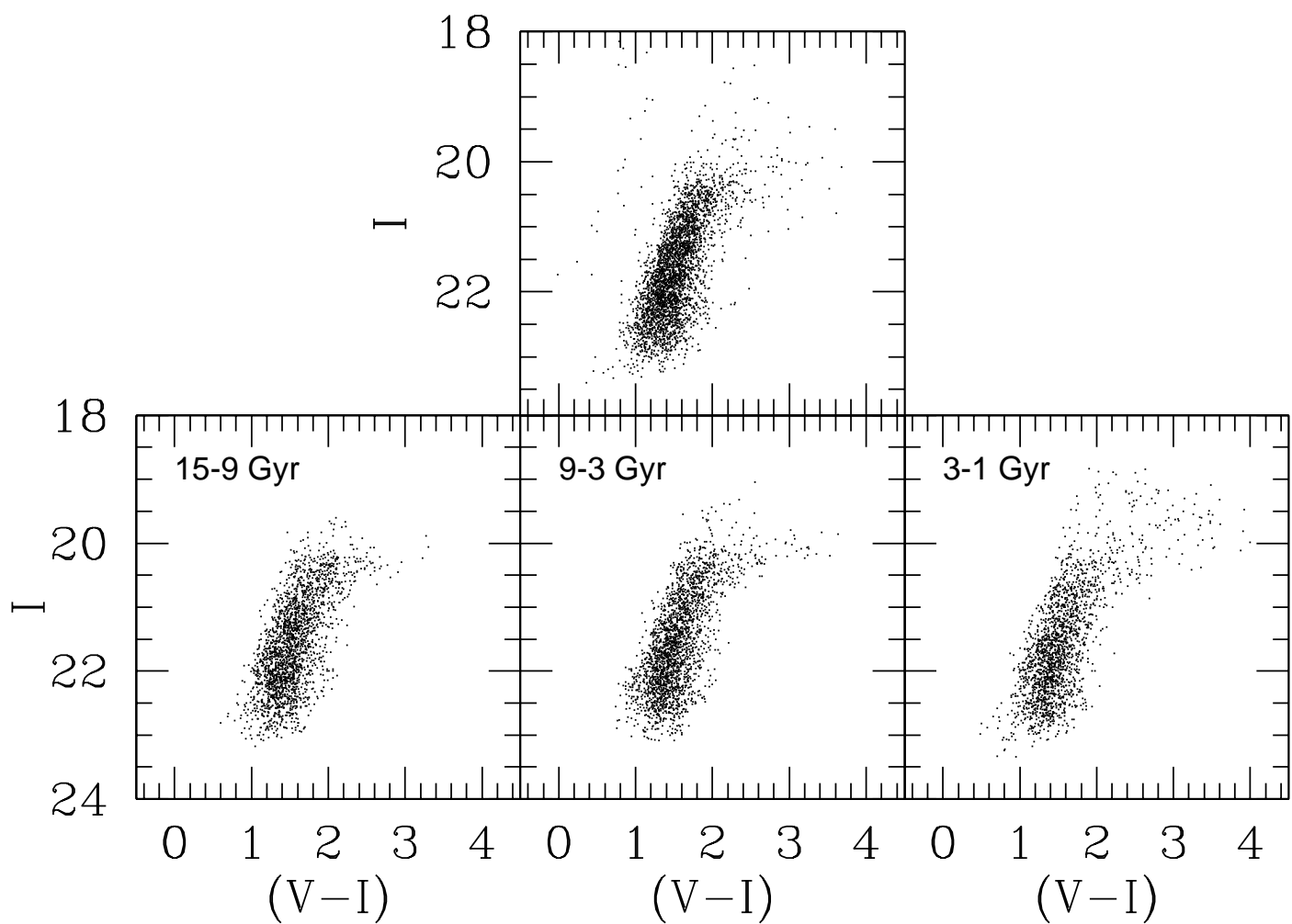
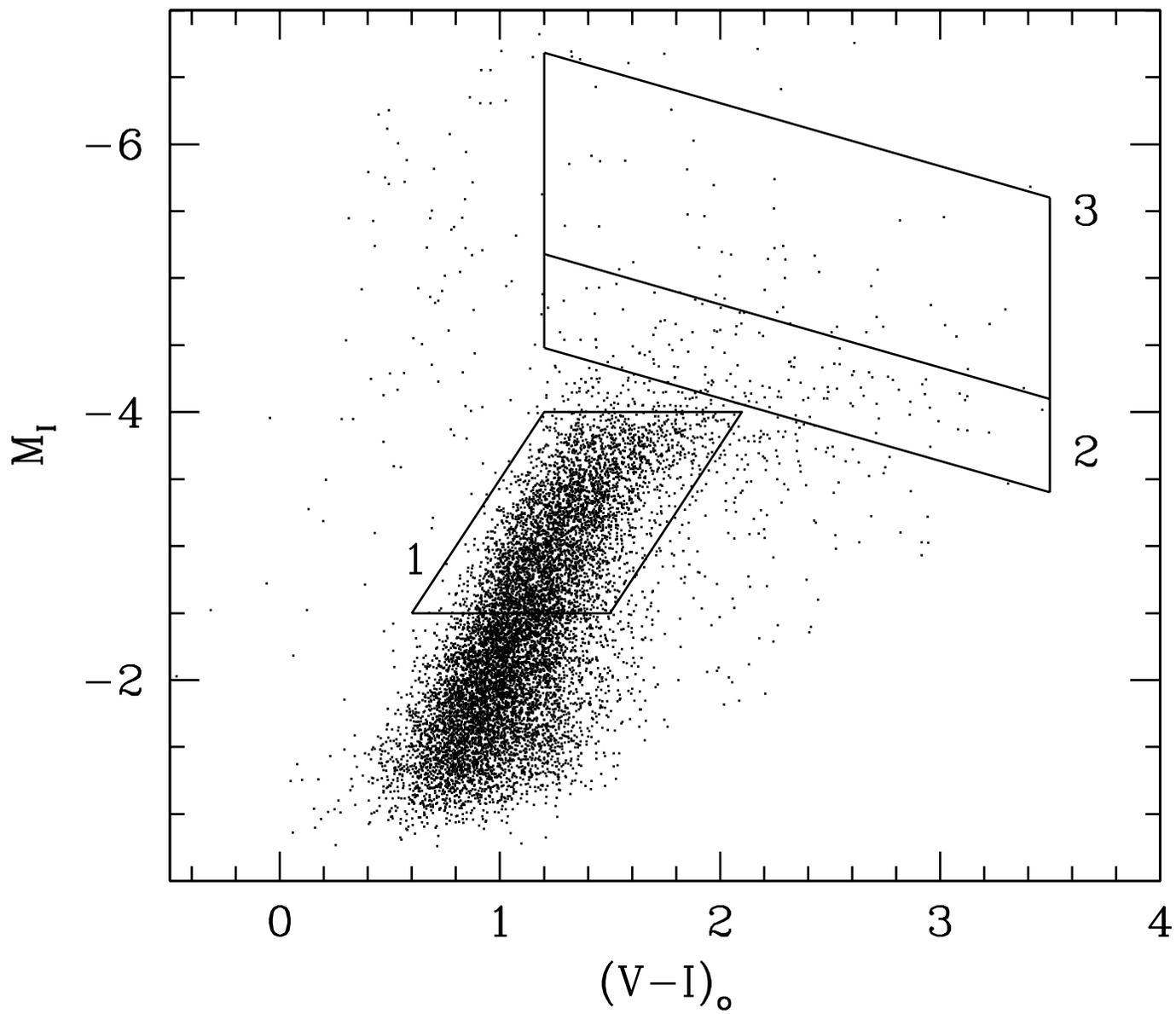
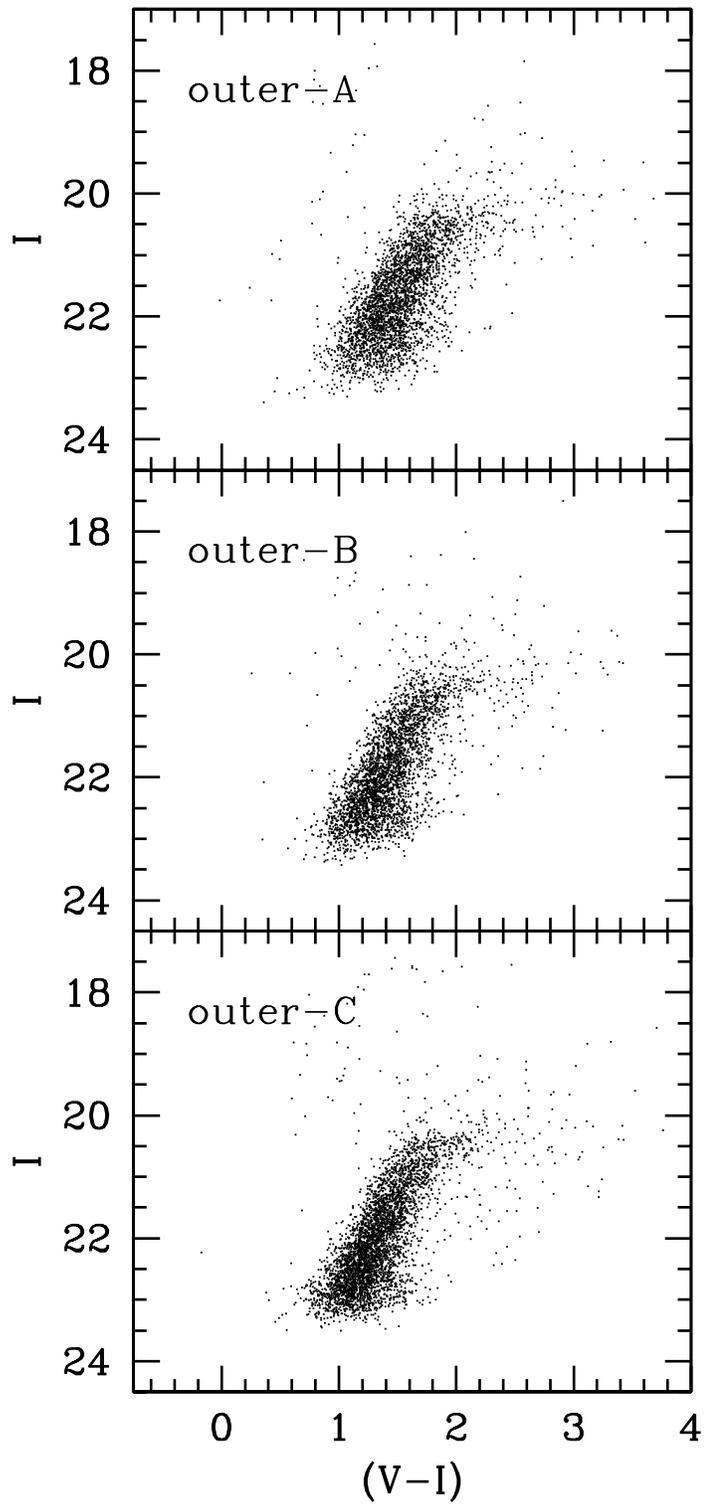
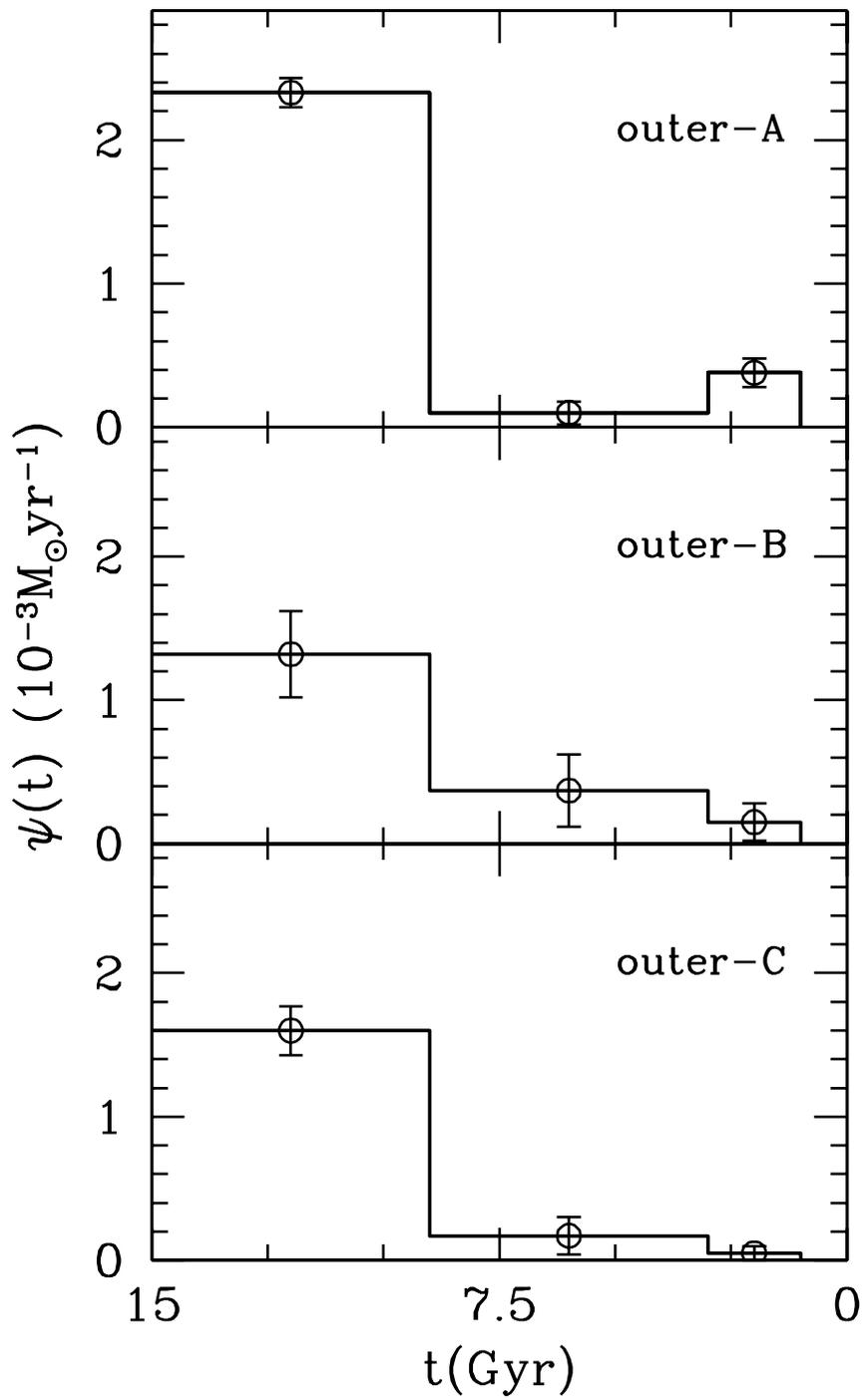
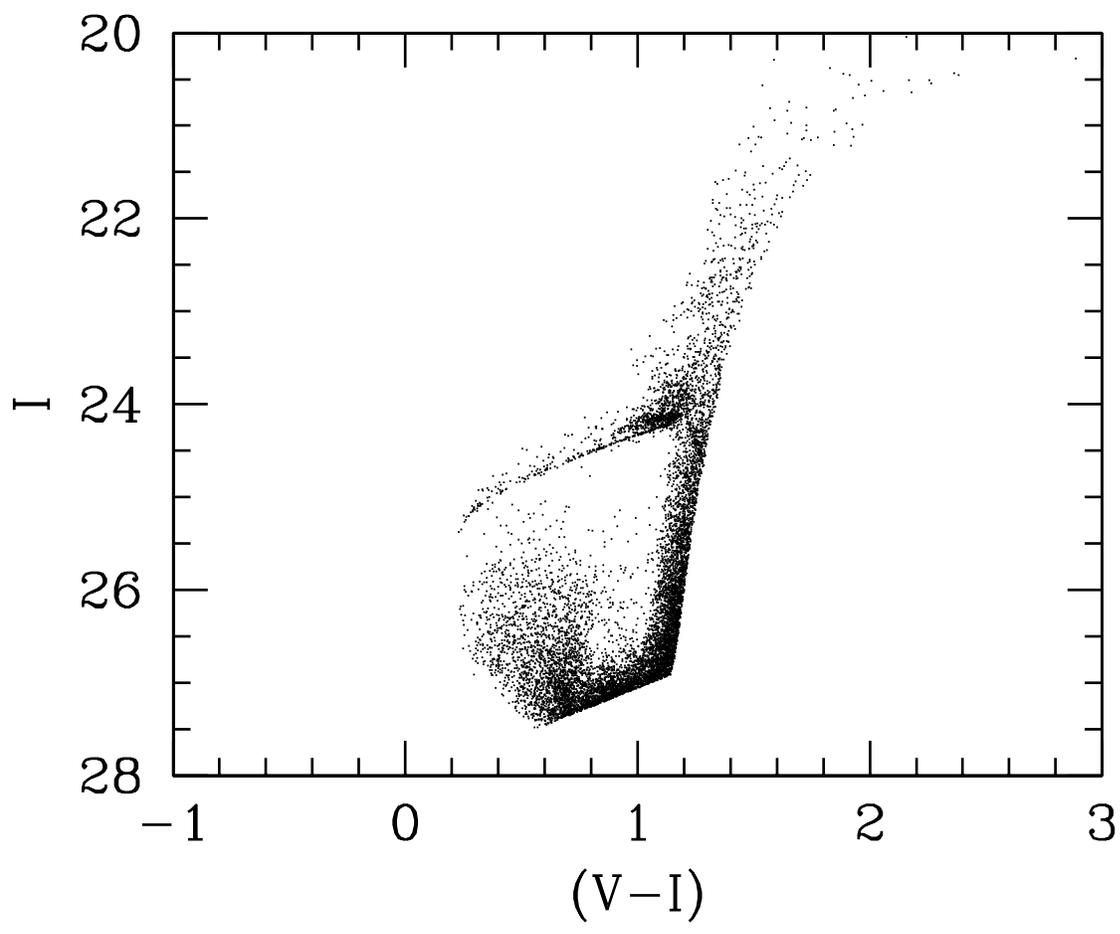
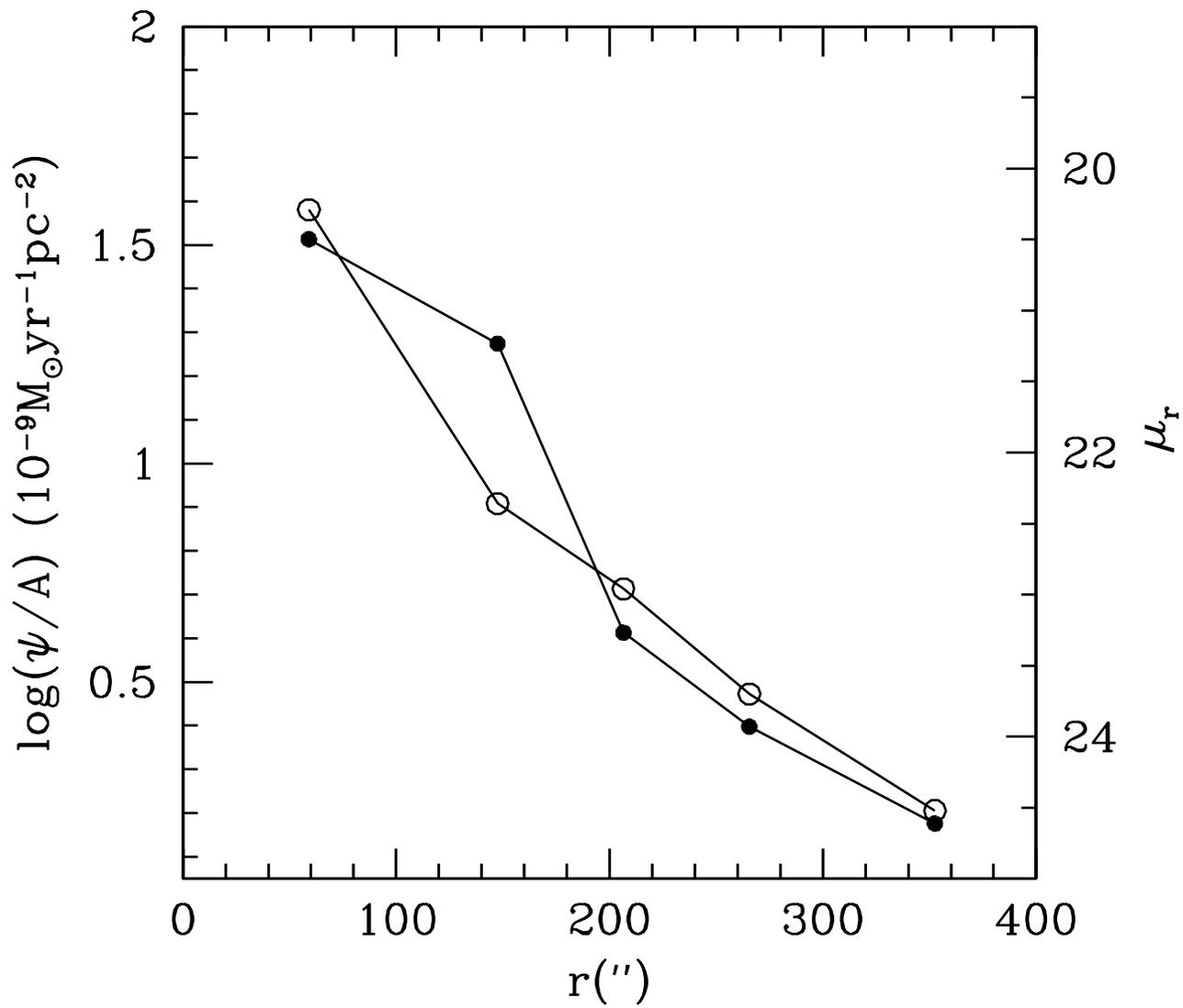
\end{document}